\documentclass[11pt]{article}
\usepackage{graphicx}
\usepackage[utf8]{inputenc}
\usepackage{charter}
\usepackage{fullpage}
\usepackage{hyperref}
\usepackage{lipsum}
\usepackage[sort&compress,numbers]{natbib}
\hypersetup{hidelinks}
\usepackage[total={6.5in,9in},top=1in,headsep=0.1in,headheight=1in]{geometry}
\usepackage{enumitem,amssymb}
\usepackage[usenames,dvipsnames]{xcolor}
\usepackage{xfrac}
\usepackage{bm}
\usepackage{array}
\usepackage{comment}
\usepackage{url}
\usepackage{hyperref}[colorlinks = true,
            linkcolor = blue,
            urlcolor  = blue,
            citecolor = blue,
            anchorcolor = blue]

\newcommand\pubnumber{The MegaMapper}
\newcommand\pubdate{September 7, 2022}

\def\Title#1{\begin{center} {\LARGE #1 } \end{center}}

\newcommand\pubblock{\rightline{\begin{tabular}{l} \pubnumber\\
         \pubdate \end{tabular}}}
\newenvironment{Abstract}{\begin{quotation} \begin{center}
                       {\bf ABSTRACT}
                       \end{center}\bigskip  }{\end{quotation}}

\def\beq{\begin{equation}}
\def\eeq#1{\label{#1}\end{equation}}
\def\eeqn{\end{equation}}

\newenvironment{Eqnarray}%
   {\arraycolsep 0.14em\begin{eqnarray}}{\end{eqnarray}}
\def\beqa{\begin{Eqnarray}}
\def\eeqa#1{\label{#1}\end{Eqnarray}}
\def\eeqan{\end{Eqnarray}}

\let\bar=\overbar

\def\lsim{\mathrel{\raise.3ex\hbox{$<$\kern-.75em\lower1ex\hbox{$\sim$}}}}
\def\gsim{\mathrel{\raise.3ex\hbox{$>$\kern-.75em\lower1ex\hbox{$\sim$}}}}

\def\del{\partial}
\def\Dslash{\not{\hbox{\kern-4pt $D$}}}
\def\dslash{\not{\hbox{\kern-2pt $\del$}}}
\def\pslash{\not{\hbox{\kern-2pt $p$}}}
\def\ETmiss{\not{\hbox{\kern-4pt $E$}}_T}

\def\Dlr{\mathrel{\raise1.5ex\hbox{$\leftrightarrow$\kern-1em\lower1.5ex\hbox{$D$}}}}

\def\MSB{{\bar{M \kern -2pt S}}}
\def\msb{{\bar{\scriptsize M \kern -1pt S}}}

\def\drb{{\bar{\scriptsize D \kern -1pt R}}}

\newcommand\apjl{{Astrophysical Journal, Letters}}%
\newcommand\apjs{{ApJS}}%
\newcommand\ao{{Appl.~Opt.}}%
\newcommand\aap{{Astronomy and Astrophysics}}%
\newcommand\snowmass{\begin{center}\rule[-0.2in]{\hsize}{0.01in}\\\rule{\hsize}{0.01in}\\
\vskip 0.1in Submitted to the  Proceedings of the US Community Study\\ 
on the Future of Particle Physics (Snowmass 2021)\\ 
\rule{\hsize}{0.01in}\\\rule[+0.2in]{\hsize}{0.01in} \end{center}}

\usepackage{authblk}

\begin{document}

\pubblock
\snowmass
\Title{The MegaMapper: A Stage-5 Spectroscopic Instrument Concept for the Study of Inflation and Dark Energy}

\bigskip 

\vspace*{-0.3cm}
\begin{center}{{\Large \textsc{Principal Authors}}}\\
\vspace*{0.3cm}
David J. Schlegel$^{1}$,
Juna A. Kollmeier $^{2,32}$ 
\vspace{0.6cm}
\end{center}

\vspace*{-0.3cm}
\begin{center}{{\Large \textsc{Team Members}}}\\
\end{center}
\noindent


\newcommand{\Amherst}{University of Massachusetts, Amherst, MA 01003 USA}
\newcommand{\ANLHEP}{HEP Division, Argonne National Laboratory, Lemont, IL 60439, USA}
\newcommand{\APC}{Laboratoire Astroparticule et Cosmologie (APC), CNRS/IN2P3, Universit\'e Paris Diderot, 10, rue Alice Domon et Léonie Duquet, 75205 Paris Cedex 13, France}
\newcommand{\ASU}{Arizona State University, Tempe, AZ  85287}
\newcommand{\BenGurion}{Department of Physics, Ben-Gurion University, Be'er Sheva 84105, Israel}
\newcommand{\BNL}{Brookhaven National Laboratory, Upton, NY 11973}
\newcommand{\Brown}{Brown University, Providence, RI 02912}
\newcommand{\Bub}{Boston University, Boston, MA 02215}
\newcommand{\BU}{Boston University, Boston, MA 02215}
\newcommand{\Buffalo}{Department of Physics, University at Buffalo, SUNY Buffalo, NY 14260 USA}
\newcommand{\Caltech}{California Institute of Technology, Pasadena, CA 91125}
\newcommand{\Cardiff}{School of Physics and Astronomy, Cardiff University, The Parade, Cardiff, CF24 3AA, UK}
\newcommand{\Carleton}{Carleton University, K1S 5B6 Ottawa, Canada}
\newcommand{\Carnegie}{The Observatories of the Carnegie Institution for Science, 813 Santa Barbara St., Pasadena, CA 91101, USA}
\newcommand{\Cavendish}{Astrophysics Group, Cavendish Laboratory, J.J.Thomson Avenue, Cambridge, CB3 0HE, UK}
\newcommand{\CCA}{Center for Computational Astrophysics, 162 5th Ave, 10010, New York, NY, USA}
\newcommand{\CITA}{Canadian Institute for Theoretical Astrophysics, 60 St. George St., Toronto Canada}
\newcommand{\CPPM}{Aix Marseille Univ, CNRS/IN2P3, CPPM, Marseille, France}

\newcommand{\CEADAP}{D\'epartement d’Astrophysique, CEA Saclay DSM/Irfu, 91191 Gif-sur-Yvette, France}
\newcommand{\CERN}{CERN, Geneva, Switzerland}
\newcommand{\CfA}{Harvard-Smithsonian Center for Astrophysics, Cambridge, MA 02138}
\newcommand{\CFT}{Center for Theoretical Physics, Polish Academy of Sciences, al. Lotnik\'{o}w 32/46, 02-668, Warsaw, Poland}
\newcommand{\Cincinnati}{University of Cincinnati, Cincinnati, OH 45221}
\newcommand{\CNRSA}{CNRS, Laboratoire d'Annecy-le-Vieux de Physique Th\'{e}orique, France}
\newcommand{\CNYang}{C.N. Yang Institute for Theoretical Physics State University of New York Stony Brook, NY 11794}
\newcommand{\CMUCosmo}{Department 
of Physics, McWilliams Center for Cosmology, Carnegie Mellon University}
\newcommand{\Columbia}{Columbia University, New York, NY 10027}
\newcommand{\Cornell}{Cornell University, Ithaca, NY 14853}
\newcommand{\CPthree}{CP3-Origins, 5230 Odense, Denmark}
\newcommand{\daa}{Department of Astronomy and Astrophysics, University of Toronto, ON, M5S3H4}
\newcommand{\damtp}{DAMTP, Centre for Mathematical Sciences, Wilberforce Road, Cambridge, UK, CB3 0WA}
\newcommand{\DESY}{DESY,  22607 Hamburg, Germany}
\newcommand{\DFI}{Departamento de F\'isica, FCFM, Universidad de Chile, Blanco Encalada 2008, Santiago, Chile}
\newcommand{\DOE}{US. Department of Energy, Germantown, MD 20874}
\newcommand{\drexel}{Drexel University, Philadelphia, PA 19104}
\newcommand{\Duke}{Duke University and Triangle Universitites Nuclear Laboratory, Durham, NC 27708}
\newcommand{\DukePhys}{Department of Physics, Duke University, Durham, NC 27708, USA}
\newcommand{\dunlap}{Dunlap Institute for Astronomy and Astrophysics, University of Toronto, ON, M5S3H4}
\newcommand{\Durham}{Department of Physics, Lower Mountjoy, South Rd, Durham DH1 3LE, United Kingdom}
\newcommand{\ED}{University of Edinburgh, EH8 9YL Edinburgh, United Kingdom}
\newcommand{\EPFL}{Institute of Physics, Laboratory of Astrophysics, Ecole Polytechnique Fédérale de Lausanne (EPFL), Observatoire de Sauverny, 1290 Versoix, Switzerland}
\newcommand{\ETH}{ETH Zurich, Institute for Particle Physics, 8093 Zurich, Switzerland}
\newcommand{\EPFLEng}{STI IMT, École Polytechnique Fédérale de Lausanne (EPFL), 1015 Lausanne, Switzerland}
\newcommand{\FNAL}{Fermi National Accelerator Laboratory, Batavia, IL 60510}
\newcommand{\FQAUB}{Dept. de F\' isica Qu\` antica i Astrof\' isica, Universitat de Barcelona, Mart\' i i Franqu\` es 1, E08028 Barcelona, Spain}
\newcommand{\FSU}{Florida State University, Tallahassee, FL 32306}
\newcommand{\Glasgow}{University of Glasgow, G12 8QQ Glasgow, United Kingdom}
\newcommand{\GRAPPA}{GRAPPA Institute, University of Amsterdam, Science Park 904, 1098 XH Amsterdam, The Netherlands}
\newcommand{\GSFC}{Goddard Space Flight Center, Greenbelt, MD 20771 USA}
\newcommand{\GWU}{George Washington University, Washington, DC 20052}
\newcommand{\Hampton}{Hampton University, Hampton, VA 23668}
\newcommand{\HarvardPhys}{Department of Physics, Harvard University, Cambridge, MA 02138, USA}
\newcommand{\Haverford}{Haverford College, 370 Lancaster Ave, Haverford PA, 19041, USA}
\newcommand{\Hawaii}{University of Hawaii, Honolulu, HI 96822}
\newcommand{\HKUST}{The Hong Kong University of Science and Technology, Hong Kong SAR, China}
\newcommand{\houston}{University of Houston, Houston, TX 77204}
\newcommand{\IAP}{Institut d'Astrophysique de Paris (IAP), CNRS \& Sorbonne University, Paris, France}
\newcommand{\IAS}{Institute for Advanced Study, Princeton, NJ 08540}
\newcommand{\IBS}{Institute for Basic Science (IBS), Daejeon 34051, Korea}
\newcommand{\ICC}{ICC, University of Barcelona, IEEC-UB, Mart\' i i Franqu\` es, 1, E08028 Barcelona, Spain}
\newcommand{\ICCD}{Institute for Computational Cosmology, Department of Physics, Durham University, South Road, Durham, DH1 3LE, UK}
\newcommand{\ICE}{Institute of Space Sciences (ICE, CSIC), Campus UAB, Carrer de Can Magrans, s/n, 08193 Barcelona, Spain}
\newcommand{\ICRR}{Institute for Cosmic Ray Resaerch, The University of Tokyo, 456 Higashi-Mozumi, Kamioka, Hida, Gifu 506-1205, Japan}
\newcommand{\ICTP}{International Centre for Theoretical Physics, Strada Costiera, 11, I-34151 Trieste, Italy}
\newcommand{\IFAE}{Institut de Fisica d’Altes Energies, The Barcelona Institute of Science and Technology, Campus UAB, 08193 Bellaterra (Barcelona), Spain}
\newcommand{\IFPU}{IFPU - Institute for Fundamental Physics of the Universe, Via Beirut 2, 34014 Trieste, Italy}
\newcommand{\IFT}{Instituto de Fisica Teorica UAM/CSIC, Universidad Autonoma de Madrid, 28049 Madrid, Spain}
\newcommand{\IFUNAM}{IFUNAM - Instituto de F\'{i}sica, Universidad Nacional Aut\'onoma de M\'etico, 04510 CDMX, M\'exico}
\newcommand{\IHEP}{Institute of High Energy Physics, Austrian Academy of Sciences, 1050 Vienna, Austria}
\newcommand{\Imperial}{Theoretical Physics, Blackett Laboratory, Imperial College, London, SW7 2AZ, U.K.}
\newcommand{\Indiana}{Indiana University, Bloomington, IN 47405}
\newcommand{\INAFOATs}{INAF - Osservatorio Astronomico di Trieste, Via G.B. Tiepolo 11, 34143 Trieste, Italy}
\newcommand{\INAFOAS}{INAF - Osservatorio di Astrofisica e Scienza dello Spazio di Bologna, via Piero Gobetti 93/3, I-40129 Bologna, Italy}
\newcommand{\INFNCag}{Istituto Nazionale di Fisica Nucleare, Sezione di Cagliari,  09126 Cagliari, Italy}
\newcommand{\INFNCat}{Istituto Nazionale di Fisica Nucleare, Sezione di Catania, 95125 Catania, Italy}
\newcommand{\INFNG}{Istituto Nazionale di Fisica Nucleare, Sezione di Genova, 16146 Genova, Italy}
\newcommand{\INFN}{INFN – National Institute for Nuclear Physics, Via Valerio 2, I-34127 Trieste, Italy}
\newcommand{\INFNLNF}{Istituto Nazionale di Fisica Nucleare, Laboratori Nazionali di Frascati, 00044 Frascati, Italy}
\newcommand{\INFNLNS}{Istituto Nazionale di Fisica Nucleare, Laboratori Nazionali del Sud, 95125 Catania, Italy}
\newcommand{\INFNN}{Istituto Nazionale di Fisica Nucleare, Sezione di Napoli, 80125 Napoli, Italy }
\newcommand{\INFNRM}{Istituto Nazionale di Fisica Nucleare, Sezione di Roma, 00185 Roma, Italy}
\newcommand{\INFNT}{Istituto Nazionale di Fisica Nucleare, Sezione di Torino, 10125, Italy }
\newcommand{\ioa}{Institute of Astronomy, University of Cambridge,Cambridge CB3 0HA, UK}
\newcommand{\IPP}{Institute for Particle Physics, BC V8W 3P6 Victoria, Canada}
\newcommand{\IPMU}{Kavli Insitute for the Physics and Mathematics of the Universe (WPI), University of Tokyo, 277-8583 Kashiwa , Japan}
\newcommand{\IPNL}{Universit\'e de Lyon, F-69622, Lyon, France; Universit\'e de Lyon 1, Villeurbanne; CNRS/IN2P3, Institut de Physique Nucl\'eaire de Lyon}
\newcommand{\IRFU}{IRFU, CEA, Universit\'e Paris-Saclay, F-91191 Gif-sur-Yvette, France}
\newcommand{\ITFA}{Institute for Theoretical Physics, University of Amsterdam, Science Park 904, 1098 XH Amsterdam, The Netherlands}
\newcommand{\IUCAA}{The Inter-University Centre for Astronomy and Astrophysics, Pune, 411007, India}
\newcommand{\Jerusalem}{Hebrew University of Jerusalem, 91904 Jerusalem, Israel}
\newcommand{\JHU}{Johns Hopkins University, Baltimore, MD 21218}
\newcommand{\JLAB}{Thomas Jefferson National Laboratory, Newport News, VA 23606}
\newcommand{\JPL}{Jet Propulsion Laboratory, California Institute of Technology, Pasadena, CA, USA}
\newcommand{\KASSI}{Korea Astronomy and Space Science Institute, Daejeon 34055, Korea}
\newcommand{\kavli}{Kavli Institute for Cosmology, Cambridge, UK, CB3 0HA}
\newcommand{\KIAS}{School of Physics, Korea Institute for Advanced Study, 85 Hoegiro, Dongdaemun-gu, Seoul 130-722, Korea}
\newcommand{\KICP}{Kavli Institute for Cosmological Physics, Chicago, IL 60637}
\newcommand{\KIPAC}{Kavli Institute for Particle Astrophysics and Cosmology, Stanford 94305}
\newcommand{\KINGS}{King's College London, WC2R 2LS London, United Kingdom}
\newcommand{\Kobe}{Kobe University, 657-8501 Kobe, Japan}
\newcommand{\KPH}{Johannes Gutenberg University, 55128 Mainz, Germany}
\newcommand{\KPMU}{University of Tokyo, 277-8583  Kashiwa , Japan}
\newcommand{\KSU}{Kansas State University, Manhattan, KS 66506}
\newcommand{\Lafayette}{Lafayette College, Easton, PA 18042}
\newcommand{\LANL}{Los Alamos National Laboratory, Los Alamos, NM 87545}
\newcommand{\LBL}{Lawrence Berkeley National Laboratory, Berkeley, CA 94720}
\newcommand{\Leiden}{Lorentz Institute, Leiden University, Niels Bohrweg 2,Leiden, NL 2333 CA, The Netherlands}
\newcommand{\Liverpool}{University of Liverpool,  L69 7ZE Liverpool , United Kingdom}
\newcommand{\LLNL}{Lawrence Livermore National Laboratory, Livermore, CA, 94550}
\newcommand{\LPC}{Universit\'e Clermont Auvergne, CNRS/IN2P3, Laboratoire de Physique de Clermont, F-63000 Clermont-Ferrand, France}
\newcommand{\LPNHE}{Sorbonne Universit\'e, Universit\'e Paris Diderot, CNRS/IN2P3, Laboratoire de Physique Nucl\'eaire et de Hautes Energies, LPNHE, 4 Place Jussieu, F-75252 Paris, France}
\newcommand{\McGill}{McGill University, Montreal, QC H3A 2T8, Canada}
\newcommand{\Melbourne}{School of Physics, The University of Melbourne, Parkville, VIC 3010, Australia}
\newcommand{\Mines}{Colorado School of Mines, Golden, CO 80401}
\newcommand{\MIT}{Massachusetts Institute of Technology, Cambridge, MA 02139}
\newcommand{\MPE}{Max-Planck-Institut f\"{u}r extraterrestrische Physik (MPE), Giessenbachstrasse 1, D-85748 Garching bei M\"unchen, Germany}
\newcommand{\MPHeidelberg}{Max Planck Institut für Astronomie, Königstuhl 17, D–69117 Heidelberg, Germany}
\newcommand{\MPIA}{Max-Planck-Institut f\"{u}r Astrophysik, Karl-Schwarzschild-Str. 1, 85741 Garching, Germany}
\newcommand{\LUPM}{Laboratoire Univers et Particules de Montpellier, Univ. Montpellier and CNRS, 34090 Montpellier, France}
\newcommand{\NAOC}{National Astronomical Observatories, Chinese Academy of Sciences, PR China}
\newcommand{\NCBJ}{National Center for Nuclear Research, Ul.Pasteura 7,Warsaw, Poland}
\newcommand{\NCU}{National Central University, Taoyuan City 32001, Taiwan (R.O.C.)}
\newcommand{\NCSU}{Physics Department, North Carolina State Universitym, 2401 Stinson Dr, Raleigh, NC 27607}
\newcommand{\ND}{University of Notre Dame,vNotre Dame, IN 46556}
\newcommand{\NIU}{Northern Illinois University, DeKalb, Illinois 60115}
\newcommand{\NMSU}{New Mexico State University, Las Cruces, NM 88003}
\newcommand{\NOAO}{NSF's NOIRLab, 950 N. Cherry Ave., Tucson, AZ 85719 USA}
\newcommand{\Northwestern}{Northwestern University, Evanston, IL 60201}
\newcommand{\Nottingham}{University of Nottingham, NG7 2RD Nottingham, United Kingdom}
\newcommand{\NWU}{Northwestern University, Evanston, IL 60208}
\newcommand{\NYU}{New York University, New York, NY 10003}
\newcommand{\OK}{ University of Oklahoma, Norman, OK 73019}
\newcommand{\ORNL}{Oak Ridge National Laboratory, Oak Ridge, TN 37831}
\newcommand{\OSU}{The Ohio State University, Columbus, OH 43212}
\newcommand{\OU}{Department of Physics and Astronomy, Ohio University, Clippinger Labs, Athens, OH 45701, USA}
\newcommand{\OskarKlein}{Oskar Klein Centre for Cosmoparticle Physics, Stockholm University, AlbaNova, Stockholm SE-106 91, Sweden}
\newcommand{\Oxford}{The University of Oxford, Oxford OX1 3RH, UK}
\newcommand{\Oxy}{Occidental College, Los Angeles, CA 90041}
\newcommand{\ParisSud}{Universit\'{e} Paris-Sud, LAL, UMR 8607, F-91898 Orsay Cedex, France \& CNRS/IN2P3, F-91405 Orsay, France}
\newcommand{\PI}{Perimeter Institute, Waterloo, Ontario N2L 2Y5, Canada}
\newcommand{\Pitt}{University of Pittsburgh and PITT PACC, Pittsburgh, PA 15260}
\newcommand{\PNNL}{Pacific Northwest National Laboratory ,Richland, WA 99352}
\newcommand{\PNPI}{Petersburg Nuclear Physics Institute, 188300 Gatchina, Russia}
\newcommand{\Port}{Institute of Cosmology \& Gravitation, University of Portsmouth, Dennis Sciama Building, Burnaby Road, Portsmouth PO1 3FX, UK}
\newcommand{\Princeton}{Princeton University, Princeton, NJ 08544}
\newcommand{\PSU}{The Pennsylvania State University, University Park, PA 16802}
\newcommand{\Purdue}{Purdue University, West Lafayette, IN 47907}
\newcommand{\PW}{Participation Worldscope, Sedona, Arizona and Hong Kong, SAR PRC}
\newcommand{\Queens}{Queen's University , K7L 3N6 Kingston, Canada}
\newcommand{\Queensland}{The University of Queensland, School of Mathematics and Physics, QLD 4072, Australia}
\newcommand{\QMUL}{Queen Mary University of London, Mile End Road, London E1 4NS, United Kingdom}
\newcommand{\RAL}{Radio Astronomy Laboratory, University of California Berkeley, Berkeley, CA 94720, USA}
\newcommand{\Rice}{Department of Physics & Astronomy, Rice University, Houston, Texas 77005, USA}
\newcommand{\RIT}{Rochester Institute of Technology}
\newcommand{\RomaS}{Dipartimento di Fisica, Universit\`{a} La Sapienza, P. le A. Moro 2, Roma, Italy}
\newcommand{\RUG}{Kapteyn Astronomical Institute, University of Groningen, P.O. Box 800, 9700 AV Groningen, The Netherlands}
\newcommand{\Rutgers}{Department of Physics and Astronomy, Rutgers, the State University of New Jersey, 136 Frelinghuysen Road, Piscataway, NJ 08854, USA}
\newcommand{\Sanford}{Sanford Underground Research Facility, Lead, SD 57754}
\newcommand{\Sassari}{Universit\`a di Sassari, 07100 Sassari,  Italy}
\newcommand{\SCIPP}{University of California at Santa Cruz, Santa Cruz, CA 95064}
\newcommand{\Sejong}{Department of Physics and Astronomy, Sejong University, Seoul, 143-747, Korea}
\newcommand{\Sheffield}{University of Sheffield, S3 7RH Sheffield, United Kingdom}
\newcommand{\SHAO}{Shanghai Astronomical Observatory (SHAO), Nandan Road 80, Shanghai 200030, China}
\newcommand{\Siena}{Siena College, Department of Physics \& Astronomy, 515 Loudon Road, Loudonville, NY 12211, USA}
\newcommand{\SISSA}{SISSA - International School for Advanced Studies, Via Bonomea 265, 34136 Trieste, Italy}
\newcommand{\SLAC}{SLAC National Accelerator Laboratory, Menlo Park, CA 94025}
\newcommand{\SMU}{Southern Methodist University, Dallas, TX 75275}
\newcommand{\SNOLAB}{SNOLAB, Lively, ON P3Y 1N2, Canada}
\newcommand{\Stanford}{Stanford University, Stanford, CA 94305}
\newcommand{\StonyBrook}{Stony Brook University, Stony Brook, NY 11794}
\newcommand{\STSCI}{Space Telescope Science Institute, Baltimore, MD 21218}
\newcommand{\SUNYA}{University at Albany SUNY, Albany, NY 12222}
\newcommand{\SussexAstronomy}{Astronomy Centre, School of Mathematical and Physical Sciences, University of Sussex, Brighton BN1 9QH, United Kingdom}
\newcommand{\Syracuse}{Syracuse University, Syracuse, NY 13244}
\newcommand{\Tamu}{Texas A\&M University, College Station, TX 77843 }
\newcommand{\Techsource}{Techsource Incorporated, Los Alamos, NM 87544}
\newcommand{\TelAviv}{Tel-Aviv University,  69978 Tel-Aviv, Israel}
\newcommand{\Temple}{Temple University, Philadelphia, PA 19122}
\newcommand{\TIFR}{Tata Institute of Fundamental Research, Homi Bhabha Road, Mumbai 400005 India}
\newcommand{\Tsinghua}{Department of Physics and Tsinghua Center for Astrophysics, Tsinghua University, Beijing 100084, P R China}
\newcommand{\TUM}{Technical University of Munich,  80333 Munich, Germany}
\newcommand{\UA}{University of Alabama, Tuscaloosa, AL 35487}
\newcommand{\UAS}{Department of Astronomy/Steward Observatory, University of Arizona, Tucson, AZ  85721}
\newcommand{\UAM}{Universidad Aut\'onoma de Madrid, 28049, Madrid, Spain}
\newcommand{\UBC}{University of British Columbia, Vancouver, BC V6T 1Z1, Canada}
\newcommand{\UCB}{Department of Astronomy, University of California Berkeley, Berkeley, CA 94720, USA}
\newcommand{\UCBP}{Department of Physics, University of California Berkeley, Berkeley, CA 94720, USA}
\newcommand{\UCBSSL}{Space Sciences Laboratory, University of California Berkeley, Berkeley, CA 94720, USA}
\newcommand{\UCBGEN}{University of California Berkeley, Berkeley, CA 94720, USA}
\newcommand{\UCD}{University of California at Davis, Davis, CA 95616}
\newcommand{\UChicago}{University of Chicago, Chicago, IL 60637}
\newcommand{\UCI}{University of California, Irvine, CA 92697}
\newcommand{\UCLA}{University of California at Los Angeles, Los Angeles,  CA 90095}
\newcommand{\UCL}{University College London, WC1E 6BT London, United Kingdom}
\newcommand{\UCR}{University of California at Riverside, Riverside, CA 92521}
\newcommand{\UCSB}{University of California at Santa Barbara, Santa Barbara, CA 93106}
\newcommand{\UCSC}{University of California at Santa Cruz, Santa Cruz, CA 95064}
\newcommand{\UCSD}{University of California San Diego, La Jolla, CA 92093}
\newcommand{\UFL}{University of Florida, Gainesville, FL 32611}
\newcommand{\UFN}{Universit\`a Federico II di Napoli, 80125 Napoli, Italy}
\newcommand{\UGTO}{Divisi\'on de Ciencias e Ingenier\'ias, Universidad de Guanajuato, Le\'on 37150, M\'exico}
\newcommand{\UKY}{University of Kentucky, Lexington, KY 40506}
\newcommand{\UMD}{University of Maryland, College Park, MD 20742
\newcommand{\UMiami}{University of Miami, Coral Gables, FL 33124}}
\newcommand{\Umich}{University of Michigan, Ann Arbor, MI 48109}
\newcommand{\UMN}{University of Minnesota, Minneapolis, MN 55455}
\newcommand{\UnB}{Instituto de F\'{i}sica, Universidade de Bras\'{i}lia, 70919-970, Bras\'{i}lia, DF, Brazil}
\newcommand{\UNC}{University of North Carolina at Chapel Hill, Chapel Hill, NC 27599}
\newcommand{\UNH}{University of New Hampshire, Durham, NH 03824}
\newcommand{\UNIPD}{Dipartimento di Fisica e Astronomia ``G. Galilei'',Universit\`a degli Studi di Padova, via Marzolo 8, I-35131, Padova, Italy}
\newcommand{\UNM}{University of New Mexico, Albuquerque, NM 87131}
\newcommand{\UNV}{University of Nevada, Reno, NV 89557}
\newcommand{\UoM}{Jodrell Bank Center for Astrophysics, School of Physics and Astronomy, University of Manchester, Oxford Road, Manchester, M13 9PL, UK}
\newcommand{\UPenn}{Department of Physics and Astronomy, University of Pennsylvania, Philadelphia, Pennsylvania 19104, USA}
\newcommand{\UR}{Department of Physics and Astronomy, University of Rochester, 500 Joseph C. Wilson Boulevard, Rochester, NY 14627, USA}
\newcommand{\UrbanaC}{Department of Physics, University of Illinois at Urbana-Champaign, Urbana, Illinois 61801, USA}
\newcommand{\USC}{The University of South Carolina, Columbia, SC 29208}
\newcommand{\USD}{The University of South Dakota, Vermillion, SD 57069}
\newcommand{\UTD}{University of Texas at Dallas, Texas 75080}
\newcommand{\Utenn}{The University of Tennessee, Knoxville, TN 37996}
\newcommand{\Utah}{University of Utah, Department of Physics and Astronomy, 115 S 1400 E, Salt Lake City, UT 84112, USA}
\newcommand{\UVA}{University of Virginia, Charlottesville, VA 22903}
\newcommand{\Uvic}{University of Victoria, BC V8P 5C2 Victoria, Canada}
\newcommand{\UWaterloo}{Department of Physics and Astronomy, University of Waterloo, 200 University Ave W, Waterloo, ON N2L 3G1, Canada}
\newcommand{\UWMadison}{Department of Physics, University of Wisconsin - Madison, Madison, WI 53706}
\newcommand{\UW}{University of Washington, Seattle 98195}
\newcommand{\UWC}{Department of Physics \& Astronomy, University of the Western Cape, Cape Town 7535, South Africa}
\newcommand{\Vanderbilt}{Physics \& Astronomy Department, Vanderbilt University, PMB 401807, 2301 Vanderbilt Place, Nashville, TN 37235}
\newcommand{\VSI}{Van Swinderen Institute for Particle Physics and Gravity, University of Groningen, Nijenborgh 4, 9747~AG~Groningen, The~Netherlands}
\newcommand{\VT}{Virginia Tech, Blacksburg, VA 24061}
\newcommand{\VUU}{Virginia Union University, Richmond, Virginia, 23220}
\newcommand{\WCA}{Centre for Astrophysics, University of Waterloo, Waterloo, Ontario N2L 3G1, Canada}
\newcommand{\Weizmann}{Weizmann Institute of Science, 76100 Rehovot, Israel}
\newcommand{\Wellesley}{Wellesley College, Wellesley, MA 02481}
\newcommand{\wiscIce}{University of Wisconsin, Madison, WI 53706}
\newcommand{\WM}{College of William and Mary, Newport News, VA 23606}
\newcommand{\WUSL}{Washington University in St Louis, St. Louis, MO 63130}
\newcommand{\WVU}{CSEE, West Virginia University, Morgantown, WV 26505, USA}
\newcommand{\WVUGWAC}{Center for Gravitational Waves and Cosmology, West Virginia University, Morgantown, WV 26505, USA}
\newcommand{\Wyoming}{Department of Physics and Astronomy, University of Wyoming, Laramie, WY 82071, USA}
\newcommand{\Yale}{Department of Physics, Yale University, New Haven, CT 06520}
\newcommand{\Brandeis}{Department of Physics, Brandeis University, Waltham, MA 02453}
Greg Aldering$^{1}$,
Stephen Bailey$^{1}$,
Charles Baltay$^{31}$,
Christopher Bebek$^{1}$,
Segev BenZvi$^{30}$,
Robert Besuner$^{1}$,
Guillermo Blanc$^{2}$,
Adam S. Bolton$^{20}$,
Ana~Bonaca$^{2}$,
Mohamed Bouri$^{4}$,
David Brooks$^{24}$,
Elizabeth Buckley-Geer$^{29}$,
Zheng Cai$^{5}$,
Jeffrey Crane$^{2}$,
Regina Demina$^{30}$,
Joseph DeRose$^{1}$,
Arjun Dey$^{20}$,
Peter Doel$^{24}$,
Xiaohui Fan$^{6}$,
Simone Ferraro$^{1}$,
Douglas Finkbeiner$^{35}$,
Andreu Font-Ribera$^{24}$,
Satya Gontcho A Gontcho$^{1}$,
Daniel Green$^{36}$,
Gaston Gutierrez$^{29}$,
Julien Guy$^{1}$, 
Henry Heetderks$^{3}$,
Dragan Huterer$^{11}$,
Leopoldo Infante$^{2}$,
Patrick Jelinsky$^{3}$,
Dionysios Karagiannis$^{34}$,
Stephen M. Kent$^{29}$,
Alex G. Kim$^{1}$,
Jean-Paul Kneib$^{8}$,
Anthony Kremin$^{1}$,
Luzius Kronig$^{8}$,
Nick Konidaris$^{2}$,
Ofer Lahav$^{24}$,
Michael L. Lampton$^{3}$,
Martin Landriau$^{1}$,
Dustin Lang$^{16}$,
Alexie Leauthaud$^{22}$,
Michael E.~Levi$^{1}$,
Michele Liguori$^{7}$,
Eric V. Linder$^{3}$,
Christophe Magneville$^{9}$,
Paul Martini$^{19}$,
Mario Mateo$^{11}$,
Patrick McDonald$^{1}$,
Christopher J. Miller$^{11}$,
John Moustakas$^{27}$,
Adam D. Myers$^{21}$,
John Mulchaey$^{2}$,
Jeffrey A. Newman$^{14}$,
Peter E. Nugent$^{1}$,
Nikhil Padmanabhan$^{31}$,
Nathalie Palanque-Delabrouille$^{9}$,
Antonella Palmese${12,1}$,
Anthony L. Piro$^{2}$,
Claire Poppett$^{3}$,
Jason X. Prochaska$^{22}$,
Anthony R. Pullen$^{18}$,
David Rabinowitz$^{31}$,
Anand Raichoor$^{1}$,
Solange Ramirez$^{2}$,
Hans-Walter Rix$^{10}$,
Ashley J. Ross$^{19}$,
Lado Samushia$^{28}$,
Emmanuel Schaan$^{1}$,
Michael Schubnell$^{11}$,
Uros Seljak$^{1,12,13}$,
Hee-Jong Seo$^{17}$,
Stephen A. Shectman$^{2}$,
Edward F. Schlafly$^{33}$,
Joseph Silber$^{1}$,
Joshua D. Simon$^{2}$,
Zachary Slepian$^{23}$,
An\v{z}e Slosar$^{37}$,
Marcelle Soares-Santos$^{15}$,
Greg Tarl{\'e}$^{11}$,
Ian Thompson$^{2}$,
Monica Valluri$^{11}$,
Risa H. Wechsler$^{25,26}$,
Martin White$^{1,13}$,
Michael J. Wilson$^{1}$,
Christophe Y\`eche$^{9}$,
Dennis Zaritsky$^{6}$,
Rongpu Zhou$^{1}$

\noindent
$^{1}$ \LBL \\
$^{2}$ \Carnegie \\
$^{3}$ \UCBSSL \\
$^{4}$ \EPFLEng \\
$^{5}$ \Tsinghua \\
$^{6}$ \UAS \\
$^{7}$ \UNIPD \\
$^{8}$ \EPFL \\
$^{9}$ \IRFU \\
$^{10}$ \MPHeidelberg \\
$^{11}$ \Umich \\
$^{12}$ \UCBP \\
$^{13}$ \UCB \\
$^{14}$ \Pitt \\
$^{15}$ \Brandeis \\
$^{16}$ \PI \\
$^{17}$ \OU \\
$^{18}$ \NYU \\
$^{19}$ \OSU \\
$^{20}$ \NOAO \\
$^{21}$ \Wyoming \\
$^{22}$ \UCSC \\
$^{23}$ \UFL \\
$^{24}$ \UCL \\
$^{25}$ \Stanford \\
$^{26}$ \SLAC \\
$^{27}$ \Siena \\
$^{28}$ \KSU \\
$^{29}$ \FNAL \\
$^{30}$ \UR \\
$^{31}$ \Yale \\
$^{32}$ \CITA\\
$^{33}$ \STSCI\\
$^{34}$ \UWC\\
$^{35}$ \CfA\\
$^{36}$ \UCSD\\
$^{37}$ \BNL\\



\medskip

 \begin{Abstract}
\noindent
In this white paper, we present the MegaMapper concept. The MegaMapper is a proposed ground-based experiment to measure Inflation parameters and Dark Energy from galaxy redshifts at $2<z<5$. In order to achieve path-breaking results with a mid-scale investment, the MegaMapper combines existing technologies for critical path elements and pushes innovative development in other design areas.  To this aim, we envision a 6.5-m Magellan-like telescope, with a newly designed wide field, coupled with DESI spectrographs, and small-pitch robots to achieve multiplexing of 26,100. 
This will match the expected achievable target density in the redshift range of interest and provide a 15x capability over the existing state-of the art, without a 15x increase in project budget.
\end{Abstract}

\section{Introduction \& Objectives}

The mechanisms driving the accelerated expansion of the Universe in its very first moments (Inflation) and at late times (Dark Energy) represent some of the most important open problems in fundamental physics, and have been the subject of several of the Snowmass Community Science White Papers \citep[e.g.,][]{Achucarro22,Amin22,Blazek22,Dawson22, Ferraro:2022cmj}.  We refer readers to those papers, and the questions therein, which have informed the development of the MegaMapper concept.  Below, we sketch in brief the concept itself and how this concept aims to address these open problems with a singular platform.  At root, the fundamental challenge for future spectroscopic experiments is mapping speed (at fixed spectroscopic fidelity).  The MegaMapper concept combines a very wide field, a large primary aperture, and a densely packed focal plane to solve the next order of magnitude in the Mapping Speed progression.  The main principles that guide the development of the MegaMapper concept are that 1) mapping speed envisioned for the next decade should aim for order-of-magnitude transformative change (see Figure 1), and 2) the development of projects of this nature should endeavor to remain ``mid-scale" as long as practical in order to efficiently benefit from technical innovation, reduce project risk, and control cost and schedule.

\subsection{Cosmology and Fundamental Physics}

 Primordial non-Gaussianity \cite{Meerburg:2019qqi} has been identified as one of the most powerful tools to study Inflation and the origin of the primordial fluctuations. Although the simplest inflationary models predict Gaussian initial conditions, large classes of models predict  deviations from Gaussianity that leave specific imprints on the galaxy power spectrum and bispectrum \cite{Meerburg:2019qqi, Alvarez:2014vva}, thus probing physics at scales inaccessible to Earth-based colliders. Distinguishing multi-field from single-field inflation requires bounds on the ``local'' non-Gaussianity parameter to be considerably better than $\sigma(f_{NL}^{\rm local}) \approx 1$, or about an order of magnitude improvement over current constraints \cite{Akrami:2019izv}. Moreover, a model independent reconstruction of the spectrum of matter perturbations will significantly improve our understanding of the primordial fluctuations, as well as any oscillation or feature in the primordial power spectrum, often affected by the physics of inflation \cite{Slosar:2019gvt, Sailer:2021yzm}. 

Our understanding of Dark Energy \cite{Slosar:2019flp} will greatly benefit from measuring the expansion and growth of fluctuations throughout cosmic history. A lever arm that extends from low redshift (Dark Energy era) to high redshifts (matter domination era) will tightly constrain large classes of theories, and test possible modifications to General Relativity. Moreover, precision measurements of the matter power spectrum will provide tight constraints on Early Dark Energy models, providing a few percent measurement out to $z \sim 10^5$ \cite{Sailer:2021yzm, Ferraro:2022cmj}.

The large volume available at $z > 2$ will increase the number of modes measured by over an order of magnitude over current experiments, opening up an uncharted territory with huge discovery potential.  Over the next decade, the Rubin Observatory's Legacy Survey of Space and Time (LSST) \citep[e.g.][]{2019ApJ...873..111I} will begin to enable hundreds of millions of tracers to be targeted for the MegaMapper.

{\bf The increased mapping speed by more than one order of magnitude, together with targets selected from the LSST imaging, will allow us to observe a high-redshift sample at $2 \lesssim z \lesssim 5$.}
The large volume available over this redshift range will allow the MegaMapper to explore primordial physics to unprecedented precision, beyond the CMB cosmic variance limit \cite{Ferraro:2022cmj}.
Dramatic improvements in the constraints on Inflation and Dark Energy \cite{Ferraro:2019uce, Sailer:2021yzm, Ferraro:2022cmj} are possible, while relaxing assumptions such as a power-law primordial power spectrum \cite{Slosar:2019gvt, dePutter:2014hza}.  LSST imaging will enable the selection of $\approx 100$ million spectroscopic targets spanning this redshift range. These targets are a combination of 
massive galaxies and highly star-forming systems, both of which can be efficiently selected from broad-band photometric surveys \cite{Wilson:2019brt}.

The MegaMapper represents a cost-effective, low-risk version of such a survey, capable of achieving the stated science goals using proven technologies. Simultaneously, experience from the pioneering SDSS suite of surveys demonstrates that such a facility would confer a tremendous opportunity for ancillary science explorations that help answer a broad array of astronomical questions from the formation of the Milky Way to the evolution and growth of galaxies.  

Possible sites for the MegaMapper include Las Campanas Observatory (Chile), Cerro Tololo Observatory (Chile), San Pedro Martir (Mexico) and Kitt Peak National Observatory (Arizona).
The Chilean sites would have full overlap with LSST imaging, whereas the northern sites would be limited to 15,000 sq.\ deg.\ of suitable extragalactic overlap in the $-30 < \delta < +30$ equatorial region.  

\begin{figure}[h]
\includegraphics[width=16cm]{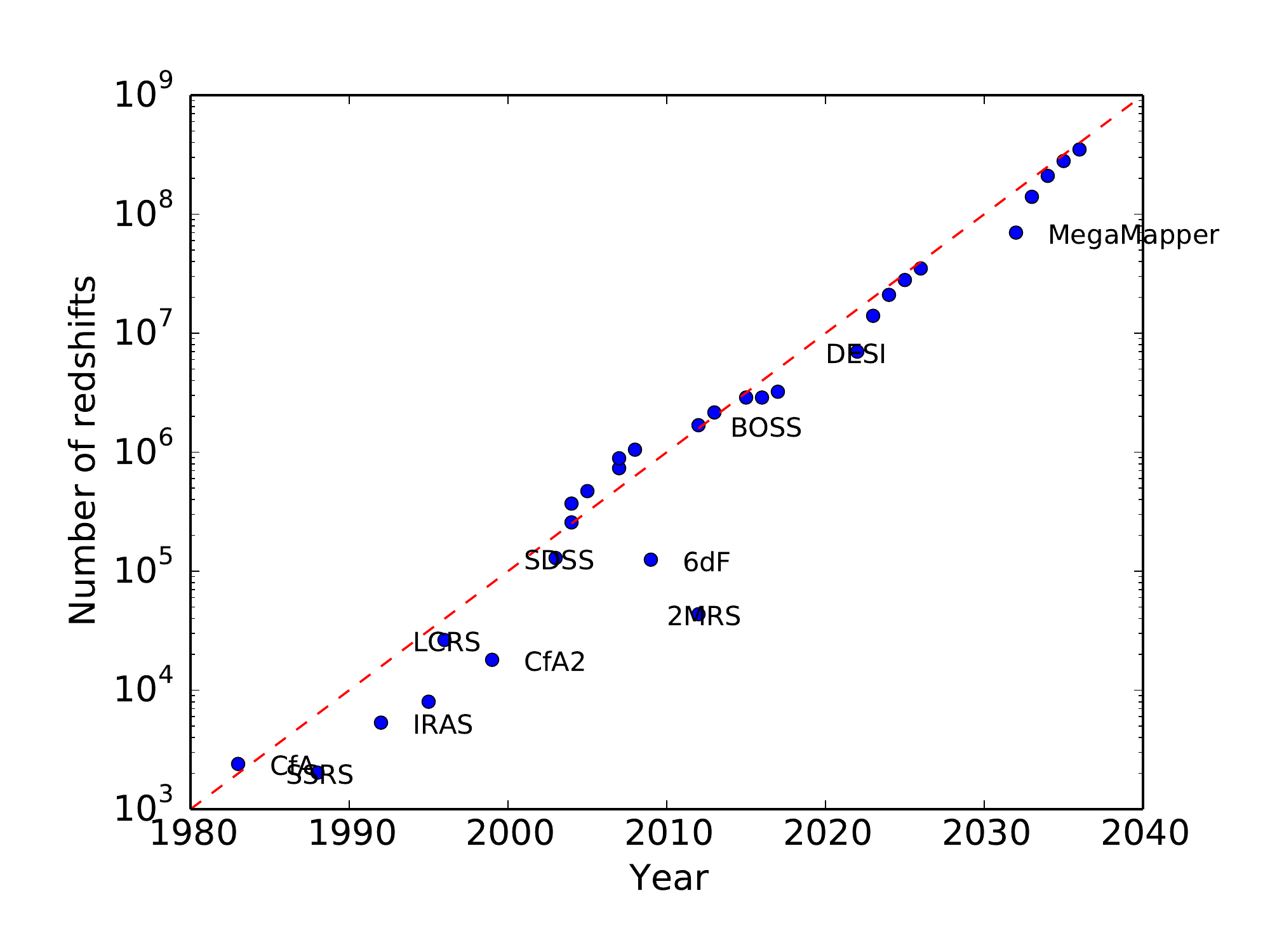}
\caption{Number of galaxy redshifts as a function of time for the largest cosmology surveys. The dotted line represents an increase of survey size by a factor of 10 every decade. Fielding the MegaMapper in ten years maintains this pace into the 2030s, and enables the Inflation and Dark Energy measures proposed in this and other white papers.}
\label{/3.ig:spectro_history}
\end{figure}

\subsection{Cosmology Science Forecasts: Inflation and Dark Energy}
In \cite{Ferraro:2019uce} we have identified two samples, a more optimistic ``idealized'' sample based on the LSST target density of Lyman-$\alpha$ emitters and Lyman-Break galaxies, and a ``fiducial'' sample, based on conservative redshift success rates and assumptions about line strengths (see \cite{Ferraro:2019uce} for the sample specifications and \cite{Wilson:2019brt} for background about selection and sample properties). 

We have found that both samples can cross the theoretical threshold $\sigma(f_{NL}^{\rm local}) \lesssim 1$, from measurement of the galaxy power spectrum on large scales, surpassing the current CMB bounds by an order of magnitude \cite{Ferraro:2019uce, Sailer:2021yzm}. Potentially large improvements are possible when including the analysis of the bispectrum. 
Improvements by a factor of two or larger over the current bounds are also expected for the equilateral and orthogonal shapes \cite{Ferraro:2019uce}. Particular care should be taken in the telescope and survey design to control systematics to allow such precise measurements of primordial non-Gaussianity.

Using a combination of Redshift-Space Distortions (RSD) and Baryon Acoustic Oscillations (BAO), the fraction of Dark Energy $\Omega_{DE}$ can be measured to better than 1\% all the way to $z \approx 4.5$ for the ``idealized'' sample, and better than 2\% up to $z \approx 5$ for the ``fiducial'' sample (see Figure 1 in \cite{Ferraro:2019uce}). Other notable improvements include a factor of two better determination of the spatial Curvature (compared to DESI + Planck), and a factor of $\gtrsim 2.5$ improvement in the Dark Energy figure of merit (Table 4 in \cite{Ferraro:2019uce}). Early Dark Energy (EDE) has been proposed as a solution to the Hubble tension \cite{Abdalla:2022yfr}. MegaMapper will be able to constrain the fraction of EDE to better than $2\%$ all the way to $z \sim 10^5$ \cite{Sailer:2021yzm}, when the Universe was only a few years old, definitely testing this hypothesis, and more generally providing percent-level expansion constraints throughout cosmic history.

\begin{table}[htb]
\caption{Survey speeds for multi-fiber spectrographs as measured by the product of the telescope clear aperture, number of fibers and losses from mirror reflections.  This speed assumes a dedicated facility, which would not be possible in all cases.  Keck/FOBOS\cite{2019arXiv190707195B}, MSE\cite{2019arXiv190707192M}, SpecTel\cite{2019arXiv190706797E} and MegaMapper\cite{2019BAAS...51g.229S} are proposed experiments.  LSSTspec\cite{2019arXiv190504669S,Blum22} is a notional number using MegaMapper positioners on the LSST focal plane, if optical design limitations could be overcome injecting f/1.2 light into fibers.
}
\label{tab_surveyspeed}
\begin{tabular}{lrrrrr@{.}l}\hline
Instrument (year)     & Primary/m$^2$ &  Nfiber  &  Reflections&  Product  & Speed vs& \ SDSS \\
          \hline
SDSS (1999)      &   3.68      &     640  &   0.9$^2$     &     1908  &    1&00 \\
BOSS (2009)     &   3.68      &    1000  &   0.9$^2$     &     2980  &    1&56 \\
DESI (2020)     &   9.5       &    5000  &   0.9$^1$     &   42,750  &   22&4 \\
PFS  (2023)     &  50         &    2400  &   0.9$^1$     &  108,000  &   56&6 \\
4MOST (2023)    &  12         &    1624  &   0.9$^2$     &      15,800  &    8&3 \\
 DESI-Upgrade (2027)     &    9.5      &    11,250  &    0.9$^1$     &   96,200  &   50&4 \\
{\bf MegaMapper}&    {\bf 28}       &    {\bf 26,100}&     {\bf 0.9}$^{\bf 2}$     &  {\bf 590,000}  &  {\bf 309}&  \\
Keck/FOBOS & 77.9       &    1800  &   0.9$^3$     &  102,000  &  53&6 \\
MSE       &  78         &    3249  &   0.9$^1$     &  228,000  &  119& \\
LSSTspec  &  35.3       &    8640  &   0.9$^3$     &  222,000  &  116& \\
SpecTel   &  87.9       &   15,000 &   0.9$^2$     & 1,070,000  &  560& \\
\hline
\end{tabular}
\end{table}

\begin{table}[h!]
\caption{Survey speeds as measured by the raw product of collecting area and field-of-view.  This is the appropriate metric for a wide-area survey with sparse targets.  Even without taking full advantage of multiplexing, the MegaMapper survey speed is competitive with larger telescopes owing to its large field-of-view.}
\label{tab_surveyarea}
\begin{tabular}{lrrrrr@{.}l}\hline
Instrument (year)      &    Primary/m$^2$  & FOV/deg$^2$ & Reflections & Product &  Speed vs& \ SDSS \\
\hline
SDSS (1999)   &     3.68     &    7.06      &  0.9$^2$   &      21.0    &    1&00 \\
BOSS (2009)   &     3.68     &    7.06      &  0.9$^2$   &      21.0    &    1&00 \\
DESI (2020)  &     9.5    &      8.04   &   0.9$^1$    &     68.7    &    3&27 \\
PFS (2023)   &    50      &      1.33   &   0.9$^1$    &      59.9    &    2&85 \\
4MOST (2023) &    12      &      4.90   &   0.9$^2$    &    58.8    &    2&80 \\
{\bf MegaMapper} &  {\bf 28}     &      {\bf 7.06}   &   {\bf 0.9}$^{\bf 2}$   &     {\bf 160.} &       {\bf 7}&{\bf{62}} \\
Keck/FOBOS  & 77.9   &     0.087  &   0.9$^3$    &     4.94    &   0&23 \\
MSE    &    78       &     1.52   &   0.9$^1$   &     107.     &    5&10 \\
LSSTspec   &    35.3    &      9.60   &   0.9$^3$   &     247.  &      11&76 \\
SpecTel    &    87.9    &      4.91   &   0.9$^2$   &     350.  &      16&65 \\
\hline
\end{tabular}
\end{table}

\subsection{Additional Science Opportunities}
The facility constructed in pursuit of the cosmology program described above also serves a broad range of additional astrophysical and cosmological objectives.  Some of these would be addressed coincident with the 5-year cosmology key project, while others could be pursued after completion of that project by the broader community.  Some of these science cases have been articulated in the National Academy of Sciences ``Astro 2020" report.  We summarize here the synergies of the MegaMapper with the recommendations of Astro2020 here.

\subsubsection{Maximizing the community investment in LSST}
A wide-field spectroscopic survey will greatly enhance the LSST science returns, as identified by several Astro 2020 science white papers \cite{Mandelbaum:2019zej, Newman:2019doi, Bechtol:2019acd}. Calibration of photometric redshifts is possible over the whole range of LSST sources through cross-correlation techniques with the spectroscopic sample. A large overlap in survey area will enable a reduction in the statistical errors to meet the stringent LSST requirements \cite{Mandelbaum:2019zej}.  The availability of a large overlapping spectroscopic sample will allow cross correlation of these galaxies with the faint LSST sources to better constrain the Intrinsic Alignment effect in weak lensing measurements.
Moreover, a combination of lensing amplitude provided by LSST, together with growth measurements through RSD can provide a powerful test of General Relativity on cosmological scales.
Finally, the MegaMapper would provide redshifts for strong gravitational lenses and type Ia supernovae discovered in LSST, allowing their cosmological interpretation. {\bf Wide-field spectroscopy is therefore a keystone for fully extracting the information content from other cosmological methodologies.}

\subsubsection{The Milky Way as a Dark Matter Experiment}

The MegaMapper, with its tremendously capable focal plane, will be a key complement to the astrophysical studies of Dark Matter \cite{Bechtol:2019acd, Drlica-Wagner:2019xan, Chakrabarti22}: by measuring the velocity dispersion of faint Milky Way satellites, the mass and density can be inferred and compared to theoretical predictions, in an environment where baryon effects are expected to be minimal. Similarly, the perturbations and gaps in the Milky Way stellar streams created by encounters with Dark Matter substructure create a characteristic velocity signal that can be measured with spectroscopy to determine the perturber mass \cite{Drlica-Wagner:2019xan}. Moreover, the merger dynamics in galaxy clusters can be studied, providing constraints on self-interacting Dark Matter \cite{Bechtol2022}.  In particular, the MegaMapper concept is already inspiring cosmologists to think about new methodological frameworks (beyond ``streams" and ``dwarf galaxies") to consider more comprehensive dark matter experiments that can be done in the Milky Way.

Additionally, of great interest in the modern era and particularly with the successful launch of the James Webb Space Telescope, the MegaMapper can enable the industrial, wide-field study of galaxy evolution to $z \sim 2$ and stellar population and kinematics and dense tomography of the intergalactic medium \cite{Najita2016}.  While it is beyond the scope of what is presented here, future instrumentation on such a facility in the post-survey period could enable a tremendous range of science from planet to galaxy formation and evolution.  The concept thus represents an {\it enabling platform} in the spirit of the ``SDSS-DESI" family of projects.

\section{Technical Overview}

\begin{figure}[htb]
\begin{center}
\includegraphics[width=14cm]{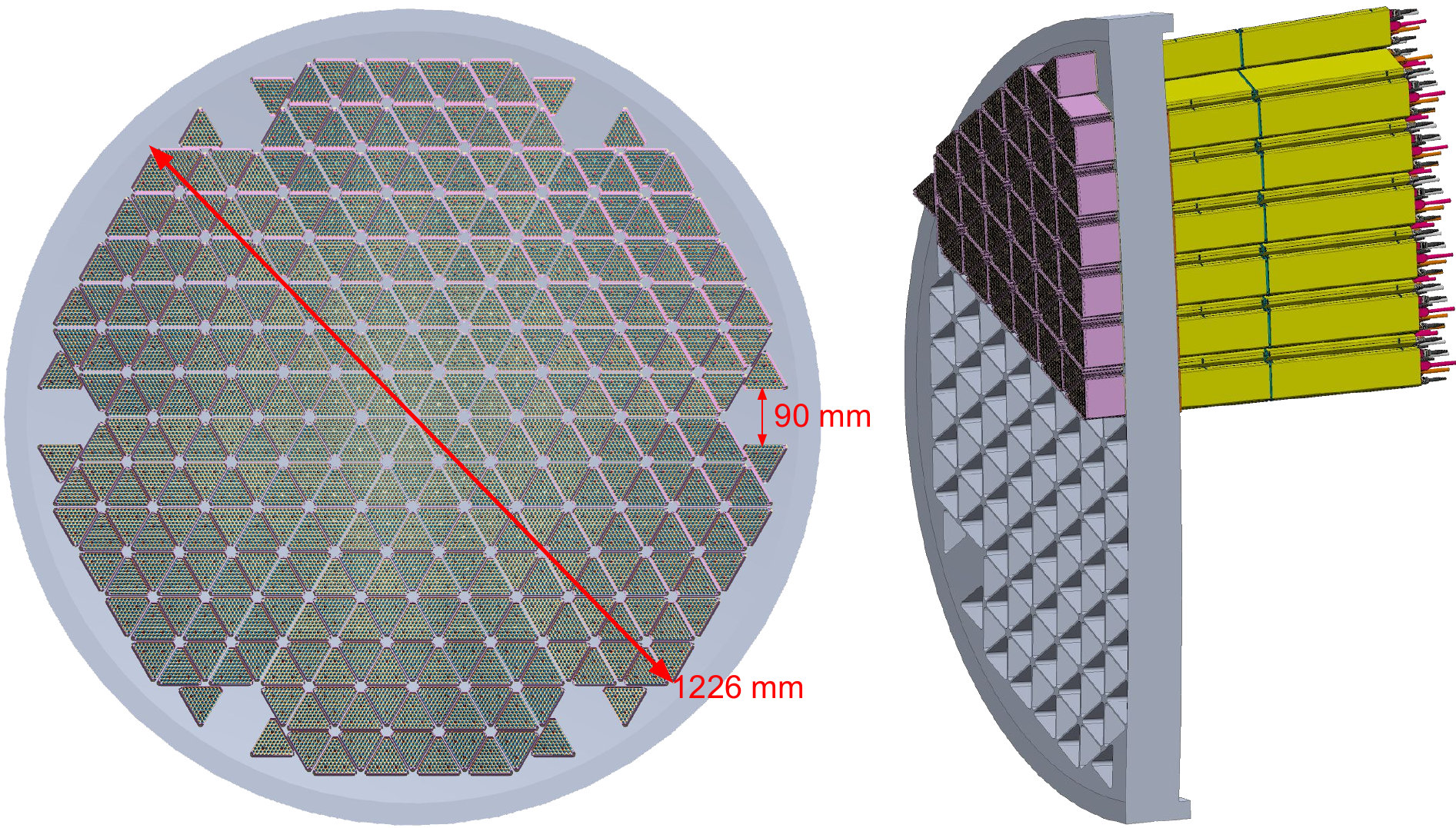}
\end{center}
\caption{Focal plane layout for 348 fiber robot rafts (with 26,100 robots) on the MegaMapper focal plane.
The triangular rafts offer mechanical stiffness, and could be invdividually inserted from the backside of the circular mounting plate.  The six unpopulated locations at the edge of the focal plane are reserved for guide/focus cameras.
}
\label{fig:mm-focal}
\end{figure}

The overall MegaMapper design seeks to optimize survey speed with a judicious choice of telescope aperture and instrument multiplexing.  A new set of optical models achieves a speed (as measured by $A \cdot \Omega$) that is difficult to match with either smaller or larger telescopes (see Table \ref{tab_surveyspeed}).  The overall telescope design is nearly identical to the existing Magellan 6.5-m telescopes, although the preferred design modifies the central hole to be larger to accommodate the wider corrected field of the new optical design.  The instrument can accommodate a multiplexing of 26,100 and would feed between 600 and 675 fibers to each DESI spectrograph.  16 of these spectrographs already exist for the DESI and SDSS-V projects (see \cite{2016arXiv161100037D} and \cite{2017arXiv171103234K}), with another 23 to 28 spectrographs required for the full capability of the Stage-5 MegaMapper instrument (See Figure~\ref{fig:mm-focal}).

\noindent
{\bf Telescope}:

The telescope concept is based on the highly successful Magellan telescope design.  This constitutes a lightweight, honeycomb structure, 6.5~m borosilicate glass primary mirror built by the University of Arizona Mirror Lab. A baseline design is considered that adopts the same optical prescription as Magellan I and II (i.e., a f/1.25 paraboloid), equipped with a 2.4 m hyperbolic secondary mirror ($\sim$70\% larger in diameter than the current f/11 Gregorian secondaries), and a 5-lens wide field corrector that provides a 3.0~deg diameter field-of-view on a Cassegrain focal plane fed at f/3.6 (see Figure \ref{fig:opticaldesign2}). An ADC is designed as part of the corrector. The first and largest element in the corrector is 1.8~m in diameter, with the other lenses also being meter sized. The large secondary and sky baffles imply a central obscuration of $\sim$20\% of the area of the primary mirror, significantly larger than the $\sim5$\% obscuration in the current Magellans but still less than either SDSS or DESI. This baseline design requires a larger central hole in the primary mirror than the 1.3~m hole in Magellan I and II, either by enlarging the central hole in an existing mirror or casting a new mirror with a custom mold.

\begin{figure}[h]
\includegraphics[width=16cm]{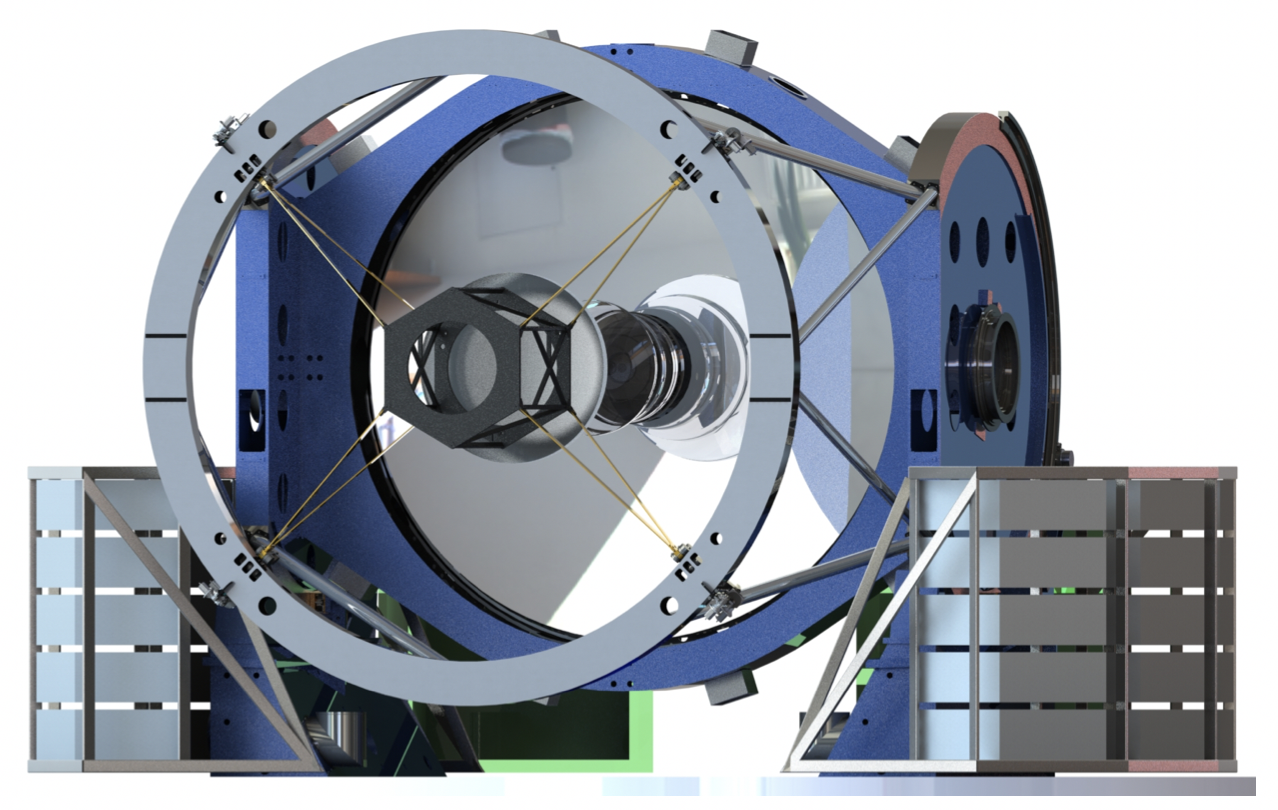}
\caption{Rendering of the Magellan-style telescope with the secondary mirror and 5-element corrector, pointed towards the horizon. The spectrographs are parked on the base  with a fiber run that is substantially shorter than the 51-meter fiber run for DESI.
}
\label{fig:jeffscad}
\end{figure}

The f/3.6 telecentric focal plane has a diameter of 1230~mm, which at a plate scale of 0.113~mm/" corresponds to a 3.0~deg diameter FOV. Figure \ref{fig:spot} presents spot diagrams at different zenith angles for the telescope's preliminary optical design described above, with $<23$~$\mu$m rms radius ($\sim 0.4$" FWHM on sky) across the full FOV, which has a maximum of 2.7\% vignetting at the field edge.
At this platescale, a 107~$\mu$m optical fiber (identical to the DESI fibers) subtends a diameter of 0.94" on the sky, which is near the optimal fiber size for point sources or compact, high-redshift galaxies in the sky-noise limit. 

The telescope mount, enclosure, and auxiliary facilities could be identical to the current Magellan design. However, particular sub-systems would be redesigned and updated.  Figure \ref{fig:jeffscad} shows the MegaMapper in the Magellan CAD.  Some subsystems would benefit from advances of technology since the Magellan 1 \& 2 builds, in particular for the sensor and control systems.

\begin{figure}[h]
\includegraphics[width=16cm]{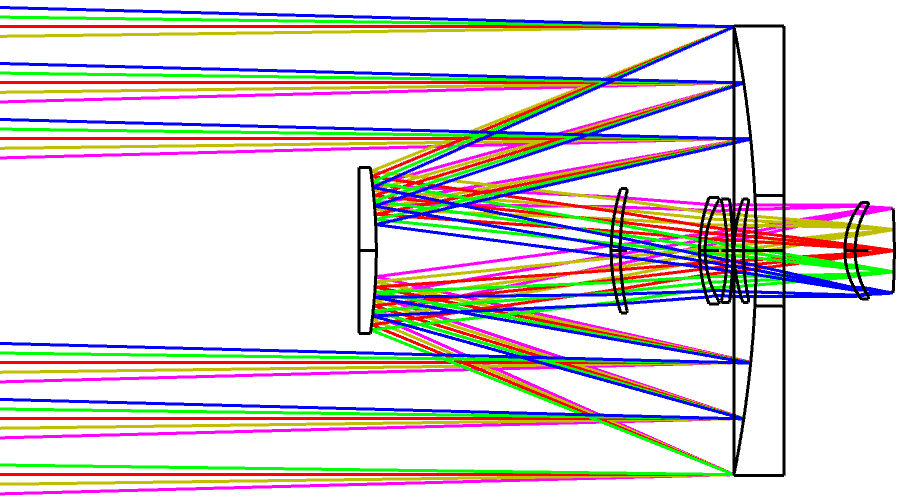}
\caption{The optical systems on the revised wide-field Magellan telescope to be used for the MegaMapper survey.  Design studies are underway on the corrector elements indicating that there are numerous vendors ready and capable of manufacturing each element of the system.  This deceptively simple diagram represents the largest objective technical risk, now largely retired since the publication of Astro2020 (Smee, private communication).}
\label{fig:opticaldesign2}
\end{figure}

\begin{figure}[htb]
\includegraphics[width=10cm, angle=-90]{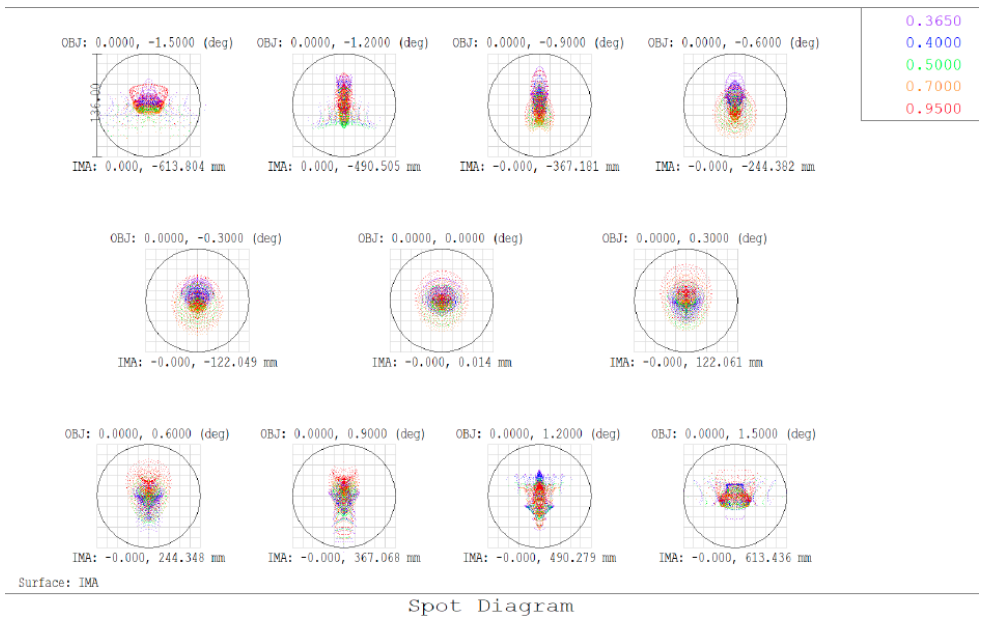}
\caption{Spot diagram for the telescope and corrector optics.  This demonstrates the excellent image quality achieved under the current optical design. (courtesy S. Smee, R. Barkhauser)}
\label{fig:spot}
\end{figure}

\begin{figure}[htb]
\includegraphics[width=16cm]{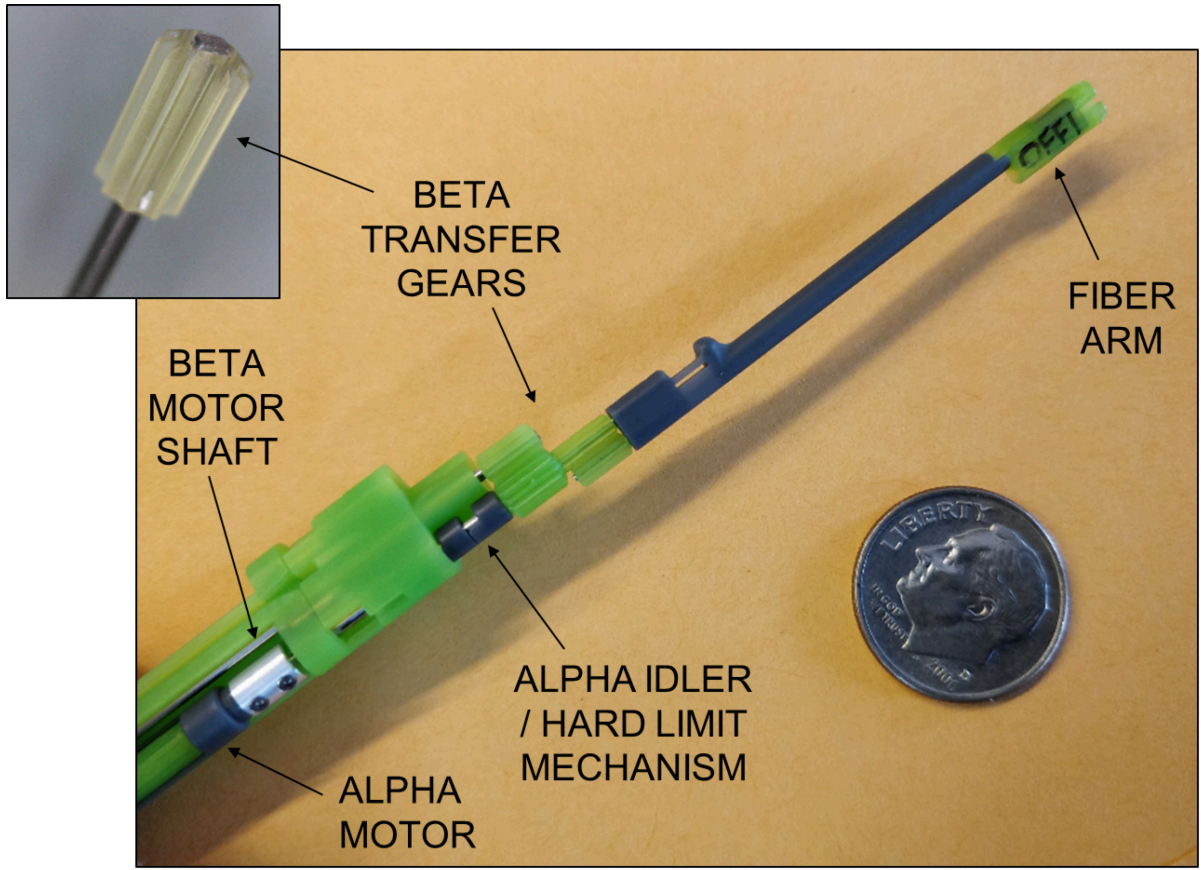}
\caption{Prototype Trillium Mark II fiber positioner under development for MegaMapper, capable of fast re-configuration of fibers to micron-scale accuracy (courtesy J. Silber).}
\label{fig:trillium}
\end{figure}

\begin{figure}[htb]
\includegraphics[width=16cm]{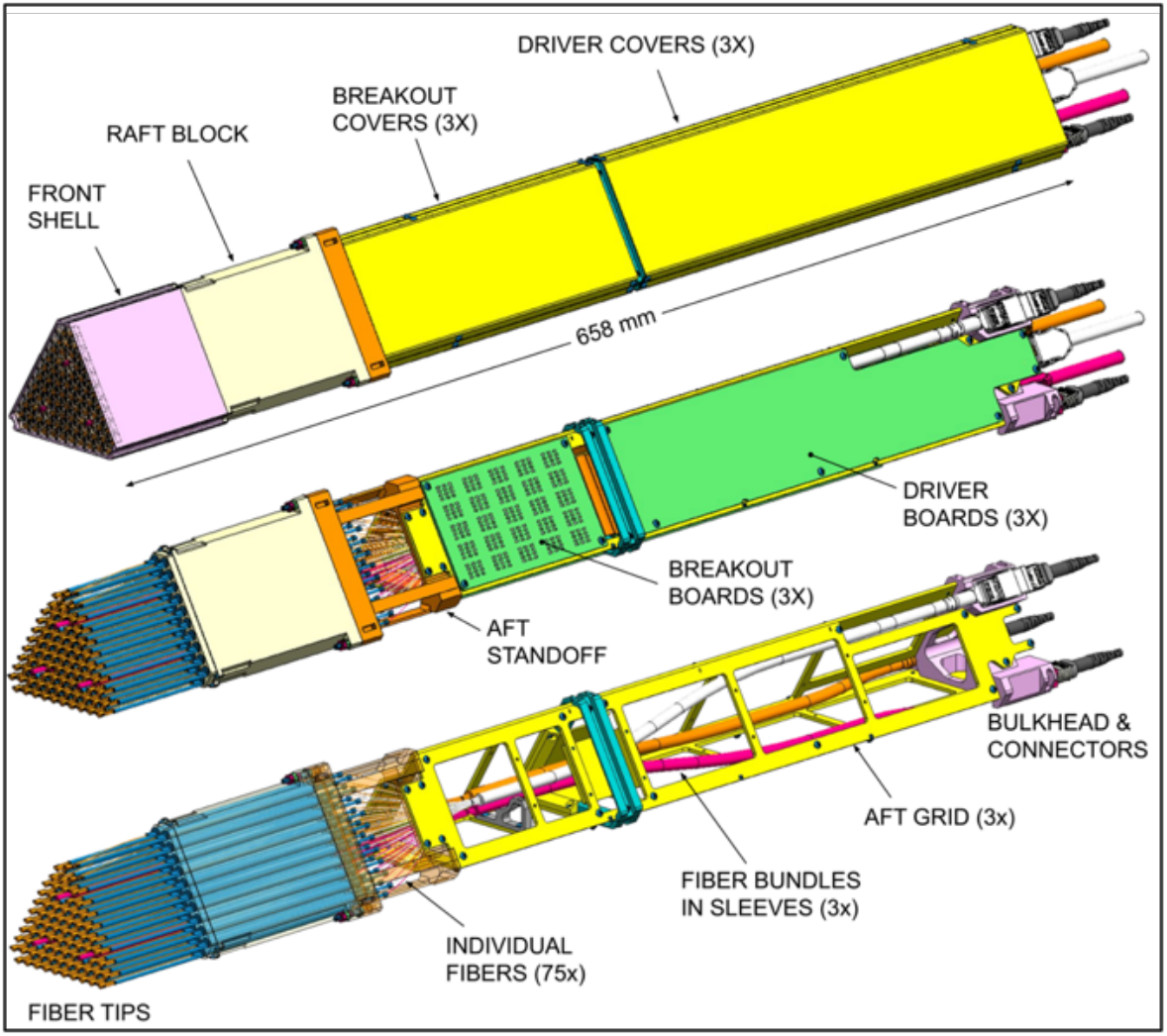}
\caption{Mechanical drawing of a 75-fiber robot raft, where the MegaMapper focal plane would consist of 348 of these rafts.  Internally, the raft is further subdivided into 3 logical and mechanical groups of 25 fibers (courtesy J. Silber).}
\label{fig:rafts}
\end{figure}

{\bf Focal Plane}:  The focal plane is physically large and accommodates 26,100 zonal fiber positioners with a  center-to-center pitch of 6.1 mm.

Each individual fiber positioner is composed of two precision, mechanical gearmotors.  Such a ``theta-phi'' motion is the basis of both the DESI and SDSS-V positioner designs.  These have the benefit of fast positioning time, high accuracy, and maintaining optical telecentricity.  DESI and SDSS-V have both completed their focal planes (of 5000 single-fiber positioners, and 1000 tri-fiber positioners respectively), demonstrating that such a system can be mass produced, reliably controlled, positioned to an accuracy of $2$~micron, and reconfigured in a time envelope of less than one minute.  

MegaMapper represents an evolution of these focal planes with smaller positioners and other design modifications based upon our experiences building the DESI and SDSS-V focal planes to facilitate mass production, integration, testing, and servicing.
Several designs have been prototyped for positioners with a center-to-center spacing of 6.2~mm, with the second generation of the Trillium positioners shown in Figure \ref{fig:trillium}.
The focal plane will be divided into 348 identical 75-fiber rafts as shown in Figure \ref{fig:rafts}, with 8 or 9 rafts feeding each spectrograph.
Each of these 75-fiber raft modules should be a complete working miniature instrument, including robotics, fibers, support structure, and electronics.
Installation and servicing of these raft modules will not require the heavy support equipment and critical lifts required for the installation of the much larger DESI focal plane petals.

{\bf Spectrographs}:  The spectrographs would be identical to those successfully built and tested for DESI and SDSS-V. These spectrographs went through extensive design and trade studies, and are optimized to measure redshifts of faint targets in the sky-noise limit.  The performance of these spectrographs has been shown to exceed their design goals in delivered optical quality and throughput.  We would choose to somewhat increase the number of fibers feeding each spectrograph from 500 to 600 or 675 by decreasing the fiber spacing at the spectrograph slit.  This is supported by the as-delivered spot quality in the as-built spectrographs.

Each spectrograph is fed by a pseudo-slit with 600-675 fibers, with dichroics dividing collimated light into three cameras.  Each camera has gratings, optics and CCDs that are optimized for its wavelength range in 360--555, 555--656, and 656--980 nm channels.  The spectral resolution runs from 2000 on the blue end (at 360~nm) increasing to a resolution of 5500 on the red end (at 980~nm) in order to work between bright sky lines.  The as-built efficiencies are 70--90\% across the full optical range.

\begin{figure}[h]
\includegraphics[width=16cm]{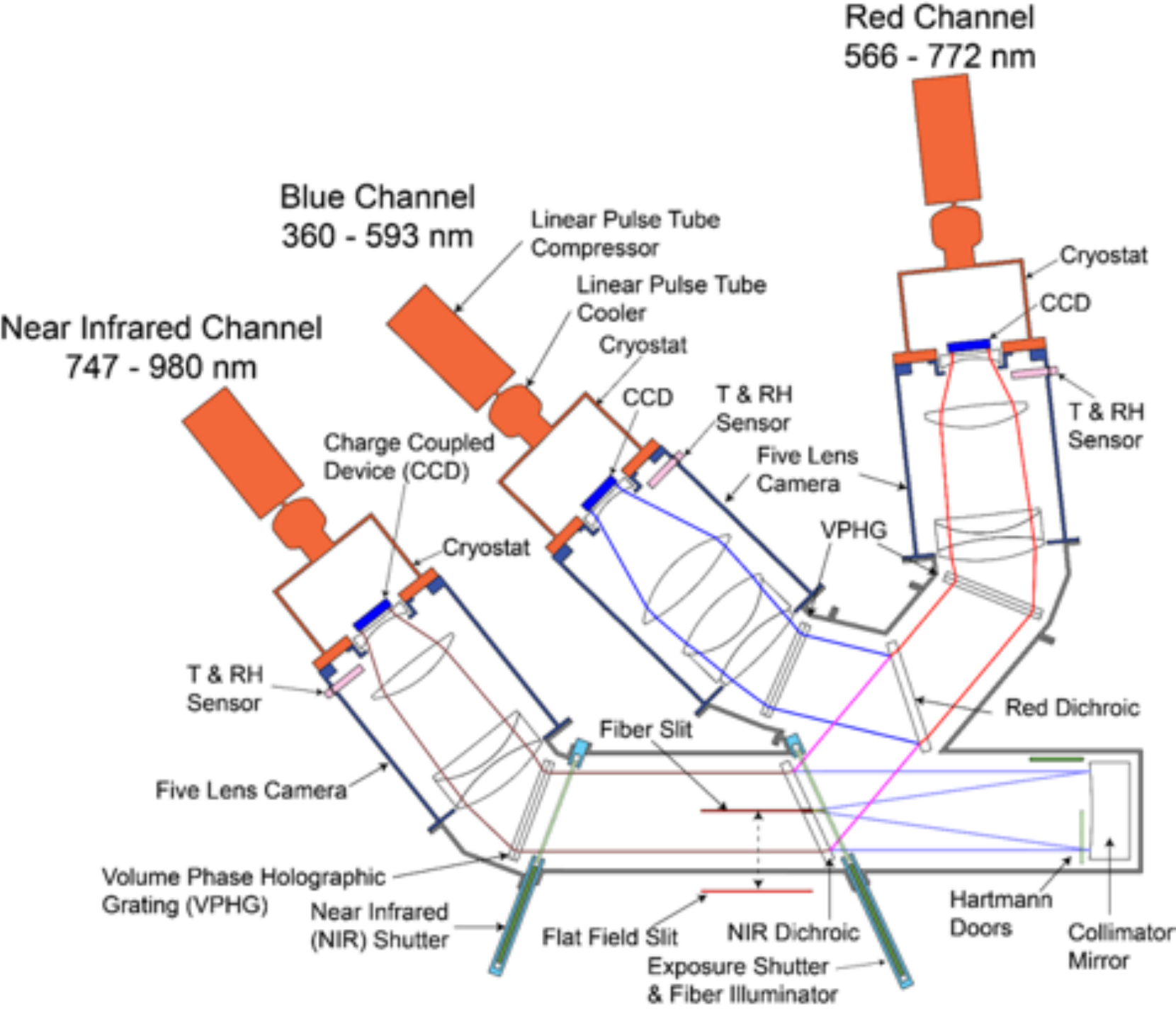}
\caption{Optical model of the DESI and SDSS-V spectrographs, which are identical systems aside from the CCD packages. As-built efficiencies are 70--90\% across the full optical spectrum.}
\label{fig:spectro_model}
\end{figure}

{\bf Data System}: The MegaMapper data systems will be a continuation of the data system as developed within the SDSS+DESI family of projects and operated on the NERSC high-performance computing platform.  The DESI data reduction from the raw pixel level to fully-calibrated spectra and redshift-fitting is state-of-the-art today, and will be maintained and updated at least through the DESI key project from 2021-2026.  Poisson-limited spectra for faint targets is achieved through a combination of stability of the spectrographs, stability of the PSF with theta-phi positioners, and a rigorous forward-modeling of the spectral extraction and sky-subtraction \cite{2010PASP..122..248B}.


\section{Conclusions}

The MegaMapper is designed to be a high-reward, low-risk, cost-effective approach to achieving a Stage-5 Spectroscopic experiment and achieving survey speed 15$\times$ faster than the current best-in-class Dark Energy Spectroscopic Instrument (DESI).  The 6.5-m primary mirror has 3$\times$ the collecting area of DESI, yet is no more challenging in the optical design for the focal plane and the spectrographs.
Our approach is to harness the power of efficient mid-scale facility construction (i.e., the Magellan telescopes) and survey design and operation (i.e., the SDSS/DESI surveys).
In combination with LSST, CMB-S4, and other cosmological probes, MegaMapper promises to explore the full range of cosmological and fundamental physics that remains essential to our understanding of the universe.







\newpage
\bibliographystyle{utphys}
\bibliography{main}

\end{document}